\documentclass[twocolumn,showpacs,preprintnumbers,amsmath,amssymb,superscriptaddress]{revtex4}

\usepackage{graphicx}
\begin{document}
\bibliographystyle{prsty}
\title{Spin and orbital magnetic moments of Fe in the $n$-type ferromagnetic semiconductor (In,Fe)As
}

\author{M.~Kobayashi}
\altaffiliation{Present address: Photon Factory, 
Institute of Materials Structure Science, High Energy Accelerator Research Organization, 
Tsukuba, Ibaraki 305-0801, Japan}
\email{masakik@post.kek.jp}
\affiliation{Department of Applied Chemistry, 
School of Engineering, University of Tokyo, 
7-3-1 Hongo, Bunkyo-ku, Tokyo 113-8656, Japan}
\affiliation{Synchrotron Radiation Research Organization, 
University of Tokyo, 1-490-2 Kouto, Sayo-cho, Tatsuno, 
Hyogo 679-5165, Japan}
\author{L.~D.~Anh}
\affiliation{Department of Electrical Engineering and Information Systems, 
University of Tokyo, 7-3-1 Hongo, Bunkyo-ku, Tokyo 113-8656, Japan}
\author{P.~N.~Hai}
\affiliation{Department of Electrical Engineering and Information Systems, 
University of Tokyo, 7-3-1 Hongo, Bunkyo-ku, Tokyo 113-8656, Japan}
\author{Y.~Takeda}
\affiliation{Synchrotron Radiation Research Unit, 
Japan Atomic Energy Agency, Sayo-gun, Hyogo 679-5148, Japan}
\author{S.~Sakamoto}
\affiliation{Department of Physics, University of Tokyo, 
7-3-1 Hongo, Bunkyo-ku, Tokyo 113-0033, Japan}
\author{T.~Kadono}
\affiliation{Department of Physics, University of Tokyo, 
7-3-1 Hongo, Bunkyo-ku, Tokyo 113-0033, Japan}
\author{T.~Okane}
\affiliation{Synchrotron Radiation Research Unit, 
Japan Atomic Energy Agency, Sayo-gun, Hyogo 679-5148, Japan}
\author{Y.~Saitoh}
\affiliation{Synchrotron Radiation Research Unit, 
Japan Atomic Energy Agency, Sayo-gun, Hyogo 679-5148, Japan}
\author{H.~Yamagami}
\affiliation{Synchrotron Radiation Research Unit, 
Japan Atomic Energy Agency, Sayo-gun, Hyogo 679-5148, Japan}
\author{Y.~Harada}
\affiliation{Synchrotron Radiation Research Organization, 
University of Tokyo, 1-490-2 Kouto, Sayo-cho, Tatsuno, 
Hyogo 679-5165, Japan}
\affiliation{Institute for Solid State Physics, The University of Tokyo, 
1-1-1 Koto, Sayo, Hyogo 679-5198, Japan}
\author{M.~Oshima}
\affiliation{Department of Applied Chemistry, 
School of Engineering, University of Tokyo, 
7-3-1 Hongo, Bunkyo-ku, Tokyo 113-8656, Japan}
\affiliation{Synchrotron Radiation Research Organization, 
University of Tokyo, 1-490-2 Kouto, Sayo-cho, Tatsuno, 
Hyogo 679-5165, Japan}
\author{M.~Tanaka}
\affiliation{Department of Electrical Engineering and Information Systems, 
University of Tokyo, 7-3-1 Hongo, Bunkyo-ku, Tokyo 113-8656, Japan}
\author{A.~Fujimori}
\affiliation{Department of Physics, University of Tokyo, 
7-3-1 Hongo, Bunkyo-ku, Tokyo 113-0033, Japan}
\date{\today}

\begin{abstract}
The electronic and magnetic properties of Fe atoms in the ferromagnetic semiconductor (In,Fe)As codoped with Be have been studied by x-ray absorption spectroscopy (XAS) and x-ray magnetic circular dichroism (XMCD) at the Fe $L_{2,3}$ edge. The XAS and XMCD spectra showed simple spectral line shapes similar to Fe metal, but the ratio of the orbital and spin magnetic moments ($M_\mathrm{orb}$/$M_\mathrm{spin}$) estimated using the XMCD sum rules was significantly larger than that of Fe metal, indicating a significant orbital moment of Fe $3d$ electrons in (In,Fe)As:Be. 
The positive value of $M_\mathrm{orb}$/$M_\mathrm{spin}$ implies that the Fe $3d$ shell is more than half-filled, which arises from the hybridization of the Fe$^{3+}$ ($d^5$) state with the charge-transfer $d^6\underline{L}$ states, where $\underline{L}$ is a ligand hole in the host valence band. 
The XMCD intensity as a function of magnetic field indicated hysteretic behavior of the superparamagnetic-like component due to discrete ferromagnetic domains. 
\end{abstract}


\maketitle

Ferromagnetic semiconductors (FMSs) are key materials for semiconductor spintronics \cite{NatMater_10_Dietl} and have attracted much attention both from the application and fundamental physics points of view. 
Charge carriers doped into the host semiconductors are considered to mediate ferromagnetic interaction between the magnetic ions, that is, so-called {\it carrier-induced ferromagnetism} \cite{RMP_06_Jungwirth}, which enables us to utilize both the charge and spin degrees of freedom of the electron for functional devices. 
Indeed, the prototypical $p$-type III-V-based FMS Ga$_{1-x}$Mn$_x$As shows ferromagnetic properties depending on carrier concentration \cite{APL_96_Ohno, PRB_98_Matsukura}, and novel spintronic functional devices using Ga$_{1-x}$Mn$_x$As have been fabricated \cite{Nature_00_Ohno, NatPhys_10_Sawicki}. However, its Curie temperature ($T_\mathrm{C}$) still remains below room temperature. 
Furthermore, in addition to the $p$-type FMSs, $n$-type FMSs are necessary for the applications of FMSs to spintronics.

Recently, Hai {\it et al}. \cite{APL_12_Hai1, APL_12_Hai2, APL_12_Hai3} have succeeded in fabricating a new $n$-type FMS In$_{1-x}$Fe$_x$As:Be (InFeAs). Here, Be atoms are located at the interstitial site under the low-temperature growth condition and act as double donors. 
One can, therefore, independently control the concentrations of magnetic dopants and electron carriers by changing the Fe and Be contents. When the electron carrier concentration is larger than $10^{19}$ cm$^{-3}$, InFeAs shows ferromagnetic properties. 
The magnetization curves show hysteresis in agreement with the magnetic field dependence of the anomalous Hall effect and magnetic circular dichroism (MCD) intensity in the visible-ultraviolet photon-energy region \cite{APL_12_Hai2}. These results indicate that the ferromagnetic property of InFeAs is intrinsic and most likely carrier-induced. 
The observation of a light electron effective mass comparable to that of the conduction electron in InAs implies that electron carriers are doped into the conduction band and that the Fermi level is located above the conduction-band minimum \cite{APL_12_Hai3}. The knowledge of the electronic structure related with the Fe ion in the InAs host is, therefore, indispensable to understand the mechanism of the ferromagnetism in InFeAs.

X-ray magnetic circular dichroism (XMCD) is a powerful tool to investigate the electronic structure of magnetic dopants in FMSs \cite{APL_00_Ohldag, PRB_05_Edmonds, PRB_05_MK, PRL_08_Takeda}. 
XMCD is defined as the difference between the x-ray absorption spectroscopy (XAS) spectra taken with circularly polarized x rays with the photon helicity parallel ($\mu^{+}$) and antiparallel ($\mu^{-}$) to the spin polarization; $\Delta \mu \equiv \mu^{+} - \mu^{-}$. Since XMCD is element-specific and sensitive only to magnetically active (ferromagnetic and paramagnetic) species, XMCD enables us to extract the electronic and magnetic properties of doped magnetic ions. 
Compared with macroscopic magnetization measurements, XMCD detects no diamagnetic signal from the substrate and has the capability to distinguish between the paramagnetic and ferromagnetic components \cite{PRL_08_Takeda}. 
In the present work, we report on the results of XMCD measurements of InFeAs, and study the role of the Fe ions in the ferromagnetism.

An In$_{1-x}$Fe$_x$As:Be ($x=0.05$) thin film with the thickness of 20 nm was grown on a InAs(001) substrate at 240 $^{\circ}$C in an ultra-high vacuum by the molecular beam epitaxy method. 
The Be concentration was $2.6 \times 10^{19}$ cm$^{-3}$. 
In order to avoid surface oxidation, the samples were covered by an As capping layer ($\sim 1$ nm) after the deposition of the In$_{1-x}$Fe$_x$As:Be layer. 
The Curie temperature $T_\mathrm{C}$ of the sample was $\sim 13$ K as determined by the Arrott plot of MCD.

X-ray absorption spectroscopy (XAS) and x-ray magnetic circular dichroism (XMCD) measurements were performed at the helical undulator beam line BL23-SU of SPring-8 \cite{JSR_98_Yokoya, AIP_04_Okamoto, JSR_12_Saitoh}. The monochromator resolution was $E/{\Delta}E \textgreater 10,000$. 
Absorption spectra ${\mu}^+$ and ${\mu}^-$ for circularly polarized x rays were obtained by reversing photon helicity at each photon energy and were recorded in the total-electron-yield mode. 
The $\mu^{+}$ and $\mu^{-}$ spectra were taken both for positive and negative applied magnetic fields and were averaged in order to eliminate spurious dichroic signals arising from the slightly different optical paths for the two circular polarizations. 
External magnetic fields were applied perpendicular to the sample surface. Backgrounds of the XAS spectra at the Fe $L_{2,3}$ edge were assumed to be hyperbolic tangent functions.

\begin{figure}[t!]
\begin{center}
\includegraphics[width=8.8cm]{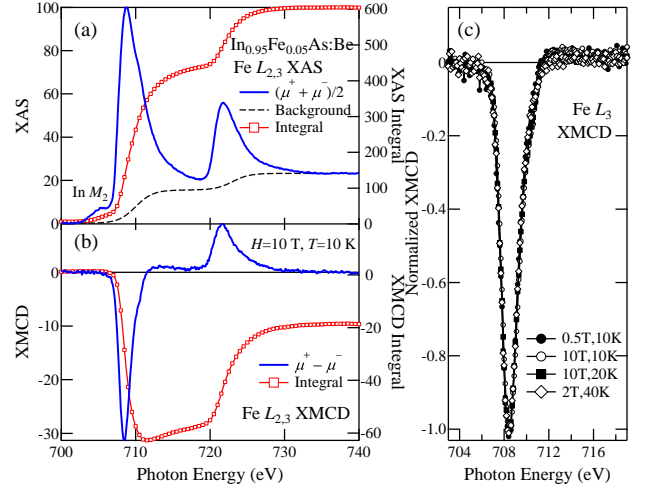}
\caption{Fe $L_{2,3}$-edge XAS and XMCD spectra of In$_{0.95}$Fe$_{0.05}$As:Be thin film. 
(a) XAS spectra [$\bar{\mu}=(\mu^{+} + \mu^{-})/2$]. 
The spectra have been normalized to the peak height ($\equiv 100$). 
(b) XMCD spectra ($\Delta \mu = \mu^{+} - \mu^{-}$) at $H = 10$ T and $T = 10$ K. 
The scale of the vertical axis is the same as panel (a). 
(c) Scaled XMCD spectra for different applied magnetic fields and temperatures. 
}
\label{IFA_XMCD}
\end{center}
\end{figure}

Figures~\ref{IFA_XMCD}(a) and ~\ref{IFA_XMCD}(b) show the Fe $L_{2,3}$ XAS and XMCD spectra of the InFeAs thin film, respectively, taken at $H = 10$ T and $T = 10$ K. 
The XAS and XMCD spectra are simple spin-orbit doublets without fine structures similar to these of Fe metal \cite{PRL_95_Chen} and Fe pnictides \cite{PRB_09_Yang}, but unlike Fe oxides \cite{JAP_09_Kataoaka}. 
The absence of clear Fe oxide signals in the spectra demonstrates that the amorphous As passivation layer protected the InFeAs surface from oxidation \cite{EL_84_Kawai}. 
The pre-edge structure around 705 eV in the XAS spectrum comes from the In $M_2$ edge. 
Figure~\ref{IFA_XMCD}(c) shows the Fe $L_3$ XMCD spectra for different $H$'s and $T$'s normalized to the main peak height. 
The almost unchanged XMCD line shapes indicate that the local electronic structure and the magnetic state of the Fe $3d$ electrons do not change with $H$ and $T$. 
However, the small shoulder structure around $h\nu = 710$ eV slightly increases with $H$, indicating that the structure may originate from a small amount of extrinsic paramagnetic oxidized Fe atoms.

The spin ($M_\mathrm{spin}$) and orbital ($M_\mathrm{orb}$) magnetic moments of Fe in (In,Fe)As in units of $\mu_\mathrm{B}$/atom are estimated by applying the XMCD sum rules \cite{PRL_92_Thole, PRL_93_Carra}: 
\begin{equation}
M_\mathrm{orb} = - \frac{2 \int_{L_{2,3}} d{\omega} \, \Delta \mu }{3 \int_{L_{2,3}} d{\omega} \, \bar{\mu}} n_h, 
\label{Morb}
\end{equation}
\begin{equation}
M_\mathrm{spin} + 7M_\mathrm{T} = - \frac{3 \int_{L_{3}} d{\omega} \, \Delta \mu - 2 \int_{L_{2,3}} d{\omega} \, \Delta \mu}{\int_{L_{2,3}} d{\omega} \, \bar{\mu} } n_h, 
\label{Mspin}
\end{equation}
to the Fe $L_{2,3}$ XAS and XMCD spectra and are listed in Table~\ref{M_XMCDsr}. Here, $n_h$ is the number of empty $3d$ states and is assumed to be 5 because of the formal valence of Fe$^{3+}$ ($3d^5$). $M_\mathrm{T}$ is the expectation value of the magnetic dipole operator and is assumed to be negligibly small because of the high symmetry of the Fe site. 
The correction factor for $M_\mathrm{spin}$ of the Fe$^{3+}$ ($d^{5}$) configuration, 0.685 \cite{PRB_09_Piamonteze}, has been applied. 
By dividing Eq.~(\ref{Morb}) to Eq.~(\ref{Mspin}), $M_\mathrm{orb}/M_\mathrm{spin}$ is given by:
\begin{equation}
M_\mathrm{orb}/M_\mathrm{spin} \approx \frac{2}{3} \, \frac{\int_{L_{2,3}} d{\omega} \, \Delta \mu}{3 \int_{L_{3}} d{\omega} \, \Delta \mu - 2 \int_{L_{2,3}} d{\omega} \Delta \mu}. 
\label{MoMs}
\end{equation}
Notably, the value of $M_\mathrm{orb}/M_\mathrm{spin}$ is independent of $n_h$ and the XAS integral. 
The finite positive value of $M_\mathrm{orb}/M_\mathrm{spin}$ reflects the finite orbital moment of a more than half-filled Fe $3d$ shell. 
Because the ratio $M_\mathrm{orb}/M_\mathrm{spin}$ of InFeAs is positive and larger than that of Fe metal \cite{PRL_95_Chen}, the electronic structure of Fe in InFeAs should be different from that of Fe in Fe metal. 
We consider that the ground state of Fe in InFeAs is basically Fe$^{3+}$ ($d^5$) (because Fe substitutes for In$^{3+}$) but that the charge-transfer $d^6\underline{L}$ configuration, where $\underline{L}$ denote a hole in the As $4p$ ligand band, is mixed into the predominant $d^5$ configuration through the Fe $3d$-As $4p$ hybridization, which indeed makes the number of $d$ electron $n_d$ ($=10 - n_h$) is a little larger than 5.

\begin{table}[!t]
\caption{Spin and orbital magnetic moments of Fe in units of Bohr magnetron in the In$_{0.95}$Fe$_{0.05}$As:Be thin film at $H = 10$ T and $T = 10$ K compared with those of Fe metal \cite{PRL_95_Chen}. 
Here, the number of $d$ electrons has been assumed to be 5. The correction factor for the Fe$^{3+}$ ion (0.685) \cite{PRB_09_Piamonteze} has been employed. 
}
\begin{tabular}{c|ccc}
\hline
\hline
\textbf{} & $M_\mathrm{spin}$ & $M_\mathrm{orb}$ & $M_\mathrm{orb}/M_\mathrm{spin}$ \\
\hline
In$_{0.95}$Fe$_{0.05}$As:Be & $1.60 \pm 0.02$ & $0.10 \pm 0.02$ & $0.065 \pm 0.014$ \\
Fe (bcc) \cite{PRL_95_Chen} & $1.95 \pm 0.08$ & $0.085 \pm 0.004$ & $0.043 \pm 0.001$ \\
\hline
\hline
\end{tabular}
\label{M_XMCDsr}
\end{table}

\begin{figure}[t!]
\begin{center}
\includegraphics[width=8.0cm]{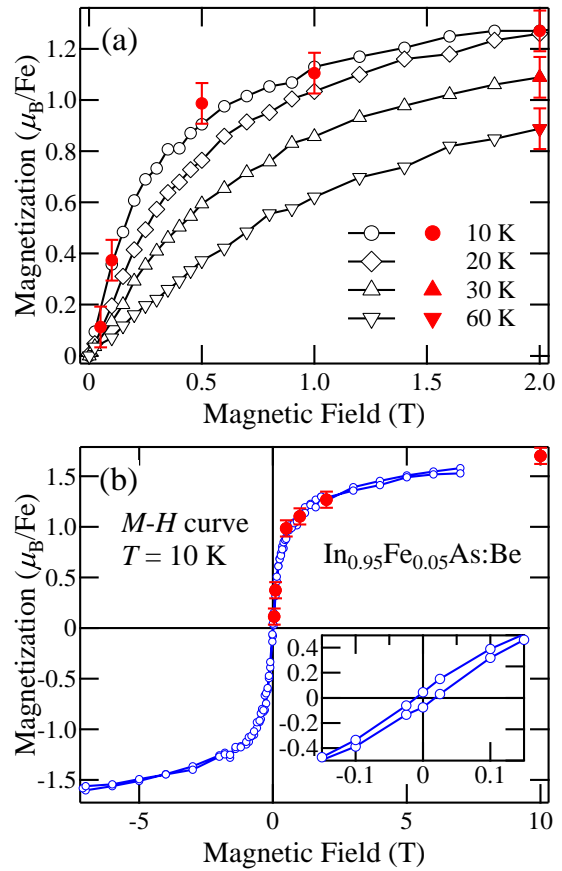}
\caption{$M$-$H$ curves deduced from the magnetic-field dependence of the Fe $L_{2,3}$ XMCD intensity. 
(a) $M$-$H$ curves taken at various temperatures. 
The XMCD peak intensity at $h\nu = 708.5$ eV (Fe $L_3$) is plotted by open symbols. 
The total magnetic moment $M = M_\mathrm{spin} + M_\mathrm{orb}$ estimated by applying the XMCD sum rules for several $H$ values is plotted by filled symbols. 
(b) $M$-$H$ curve taken at $T=10$ K for both positive and negative magnetic fields. The inset shows an enlarged plot around the zero magnetic fields. 
}
\label{IFA_MH}
\end{center}
\end{figure}

Figure~\ref{IFA_MH}(a) shows the $H$ dependence of the XMCD intensity at $h\nu = 708.5$ eV, namely, at the peak position of $L_3$ XMCD at various temperatures. 
Since the line shape of the XMCD spectrum remains nearly unchanged with varying $H$ and $T$ as shown in Fig.~\ref{IFA_XMCD}(c), the XMCD peak intensity should be proportional to the total magnetic moment of Fe $M$ (=$M_\mathrm{spin}+M_\mathrm{orb}$) or the magnetization. 
In Fig.~\ref{IFA_MH}(a), therefore, we have converted the vertical axis from the XMCD peak intensity to $M$. At several $H$ values, the XMCD sum rules have been applied to the entire spectra and the deduced $M$ values are plotted by filled symbols. 
With increasing $T$, $M$ decreases and the susceptibility, i.e., the slope of the $M$-$H$ curves around $H=0$, diminishes. 
The $M$-$H$ curves are concave, suggesting globally superparamagnetic-like behavior. 
In the previous studies \cite{APL_12_Hai2}, such superparamagnetic-like hysteresis is attributed to the existence of discrete ferromagnetic domains due to the spatial fluctuations of electron concentration $n_e$; areas with high $n_e$ become ferromagnetic, while areas with low $n_e$ remain paramagnetic. At a zero magnetic field, the magnetization is minimized to reduce the total static magnetic energy due to dipolar interactions between ferromagnetic domains, which explains the observed small remanent magnetization. 
Note that this behavior is significantly different from the superparamagnetism due to thermal fluctuations in conventional nanocluster systems. 
Indeed, the coexistence of ferromagnetic and paramagnetic domains on a 10 $\mu$m scale in InFeAs has been confirmed by magneto-optical imaging \cite{APL_12_Hai2}.

To further confirm this picture, we measured an $M$-$H$ curve at $T=10$ K for magnetic fields ranging from positive to negative directions, as shown in Figure~\ref{IFA_MH}(b). 
Since the line shape of the XMCD spectra is independent of temperature, the electronic structure of the intrinsic paramagnetic Fe components is indistinguishable from that of the ferromagnetic one. 
The coexistence of ferromagnetic and paramagnetic regions is evidenced by two characteristics. First, the saturated magnetic moment per Fe atom of the ferromagnetic component is only 1.2 $\mu_\mathrm{B}$, which is much smaller than the expected $\sim5$ $\mu_\mathrm{B}$ if all Fe atoms contributed to ferromagnetism. This indicates that there must be regions that remain paramagnetic. Second, the $M$-$H$ curve is not saturated and has a linear component even at 7 Tesla, which indicates the contribution of paramagnetic areas. Nevertheless, the sample is not superparamagnetic, since clear coercivity of $\sim 40$ Oe and non-zero remanent magnetization are observed, as shown in the inset of Fig.~\ref{IFA_MH}(b). 
This is consistent with those obtained from the SQUID and MCD measurements \cite{APL_12_Hai2} and gives evidence for the ferromagnetic behavior of the InFeAs with $T_\mathrm{C}=13$ K.

In conclusion, we have performed an XMCD study on InFeAs thin film to investigate the electronic structure of the Fe ions related to the magnetism. 
The line shape of the XMCD is unchanged with magnetic fields and temperatures, the latter of which indicates that the XMCD signal from doped Fe atoms is intrinsic. 
The ratio $M_\mathrm{orb}/M_\mathrm{spin}$ is positive and larger than that of Fe metal, indicating that the electronic structure of the Fe ions in InFeAs is different from that of Fe metal. 
The positive $M_\mathrm{orb}/M_\mathrm{spin}$ is explained by the charge transfer from the ligand to Fe $3d$ orbitals through the hybridization between the Fe $3d$ and As $4p$ ligand states. 
The XMCD intensity as a function of magnetic field shows hysteretic behavior of the superparamagnetic-like component due to discrete magnetic domains in InFeAs:Be. 
We suggest that, in order to improve the magnetic characteristics of InFeAs, homogenous electron doping is necessary, which may be obtained if group VI atoms are used as donors in InFeAs.

This work was supported by Grants-in-Aid for Scientific Research (S22224005 and 23000010) from the Japan Society for the Promotion of Science (JSPS), Japan. 
The experiment at SPring-8 was approved by the Japan Synchrotron Radiation Research Institute (JASRI) Proposal Review Committee (Proposal No.2012B3823). 
MK acknowledges financial support from JSPS.

\end{document}